%% file: EDS_nov_11.tex
\documentclass[12pt]{article}
\usepackage{graphicx}
\usepackage{amssymb, graphics, setspace}
\usepackage{pgfplots}
\usepackage{textcomp}
\usepackage[caption=false]{subfig}

\usepackage{amsmath}
\usepackage{commath}

\usepackage{graphicx,bm}
\usepackage{url}
\usepackage{float}
\usepackage{textcomp}
\usepackage{verbatim}
\def\napoli{Uzhgorod National University, \\14, Universytets'ka str.,  
	Uzhgorod, 88000, Ukraine \\istvanszanyi.phys@gmail.com}
\def\support{\footnote{Work was supported by HDFU (UMDSZ).}}

\def\Title#1{\begin{center} {\Large #1 } \end{center}}
\def\Author#1{\begin{center}{ \sc #1} \end{center}}
\def\Address#1{\begin{center}{ \it #1} \end{center}}

\newenvironment{Abstract}{\begin{quotation}  }{\end{quotation}}
\newenvironment{Presented}{\begin{quotation} \begin{center} 
             PRESENTED AT\end{center}\bigskip 
      \begin{center}\begin{large}}{\end{large}\end{center} \end{quotation}}
\def\Acknowledgements{\bigskip  \bigskip \begin{center} \begin{large}
             \bf ACKNOWLEDGEMENTS \end{large}\end{center}}

\input econfmacros.tex

\begin{document}
\begin{titlepage}

\vfill
\Title{Structures in the high-energy proton-proton diffraction cone}
\vfill
\Author{Istv\'an Szanyi\support}
\Address{\napoli}
\vfill
\begin{Abstract}
The otherwise exponential high-energy proton-proton diffraction cone has two prominent structures, which are namely the "break" staying fixed around $t \approx -0.1$ GeV$^2$ and the "dip" moving with increasing energy logarithmically towards smaller $|t|$ values. While at the ISR the two structures are separated by a distance of about $1$~GeV$^2$, at the LHC the dip comes close to the periphery of the "break", thus affecting its parametrization. The Regge theory gives some opportunity to describe these phenomena which have still disputable and quite different physics.\footnote{Talk based on recent pepers \cite{RPM,JSZ2}} 
\end{Abstract}
\vfill
\begin{Presented}
Presented at EDS Blois 2017, Prague, \\ Czech Republic, June 26-30, 2017
\end{Presented}
\vfill
\end{titlepage}
\def\thefootnote{\fnsymbol{footnote}}
\setcounter{footnote}{0}

\section{Introduction}
	
The diffraction cone of high-energy elastic hadron scattering deviates from a purely exponential due to two structures clearly visible in proton-proton scattering, namely a "break" (in fact, a smooth concave curvature) near $t=-0.1$ GeV$^2$, whose position is independent of energy) and the prominent "dip" - diffraction minimum moving slowly (logarithmically) with $s$ towards smaller values of $|t|$, where $s$ and $t$ are the  Mandelstam variables. While in the ISR energy region, $23.5\leq \sqrt{s}\leq 62.5$ GeV, the dip is known to be located near $-t=1.4$ GeV$^2$, at the LHC, $\sqrt {s}=7,$ TeV it was found \cite{TOTEMdip} near $t\approx-0.5$ GeV$^2$. Physics of the two phenomena is quite different and still disputable: while the "break" may be related to the two-pion threshold required by $t$-channel unitarity \cite{LNC}, the dip (diffraction minimum) is a consequence of $s$-channel unitarity or absorption corrections to the scattering amplitude \cite{Reviews}. 

\begin{figure}[H] 
	\centering
	\subfloat[\label{Fig:1a}]{%
		\includegraphics[width=0.41\textwidth]{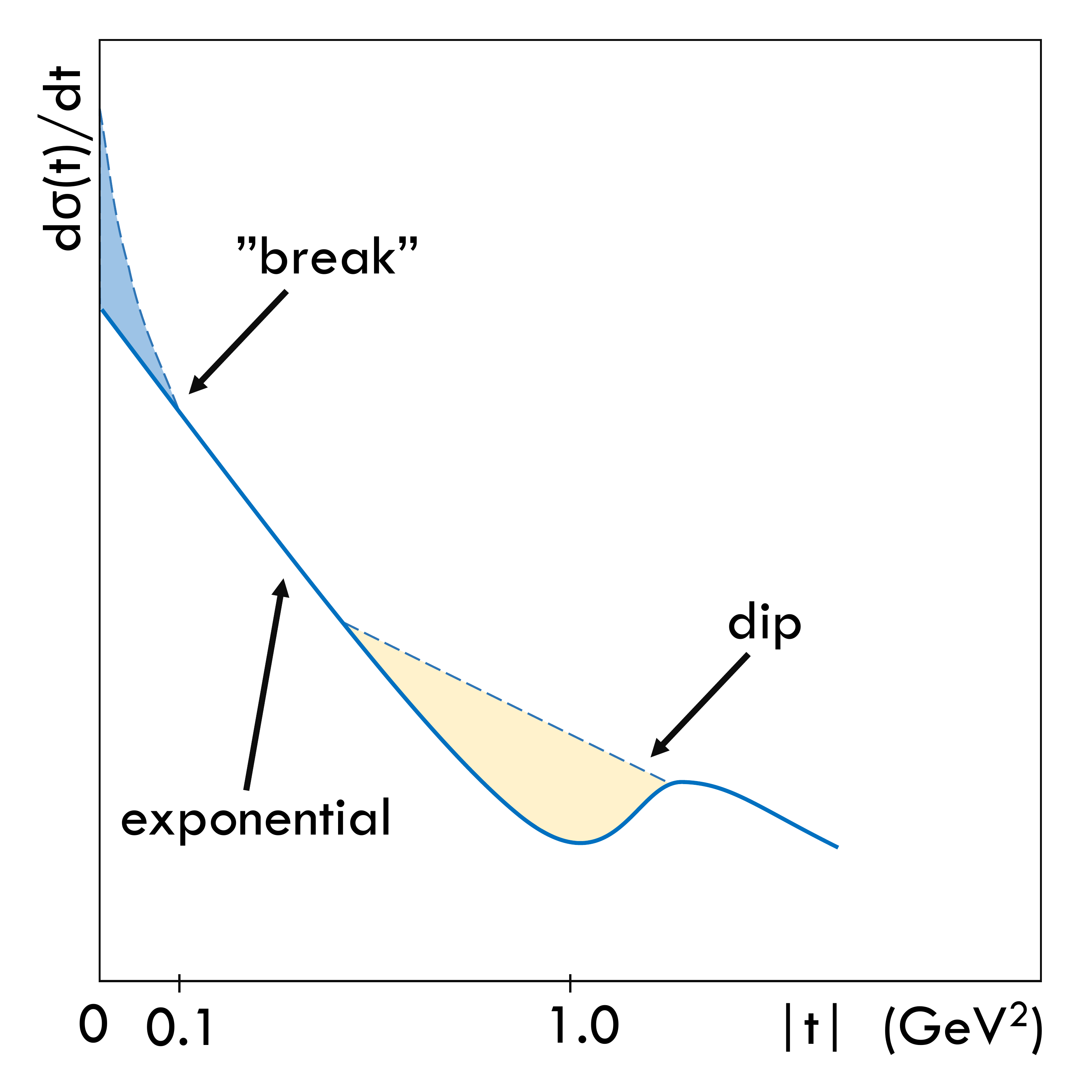}%
	}\quad
	\subfloat[\label{Fig:1b}]{%
		\includegraphics[width=0.41\textwidth]{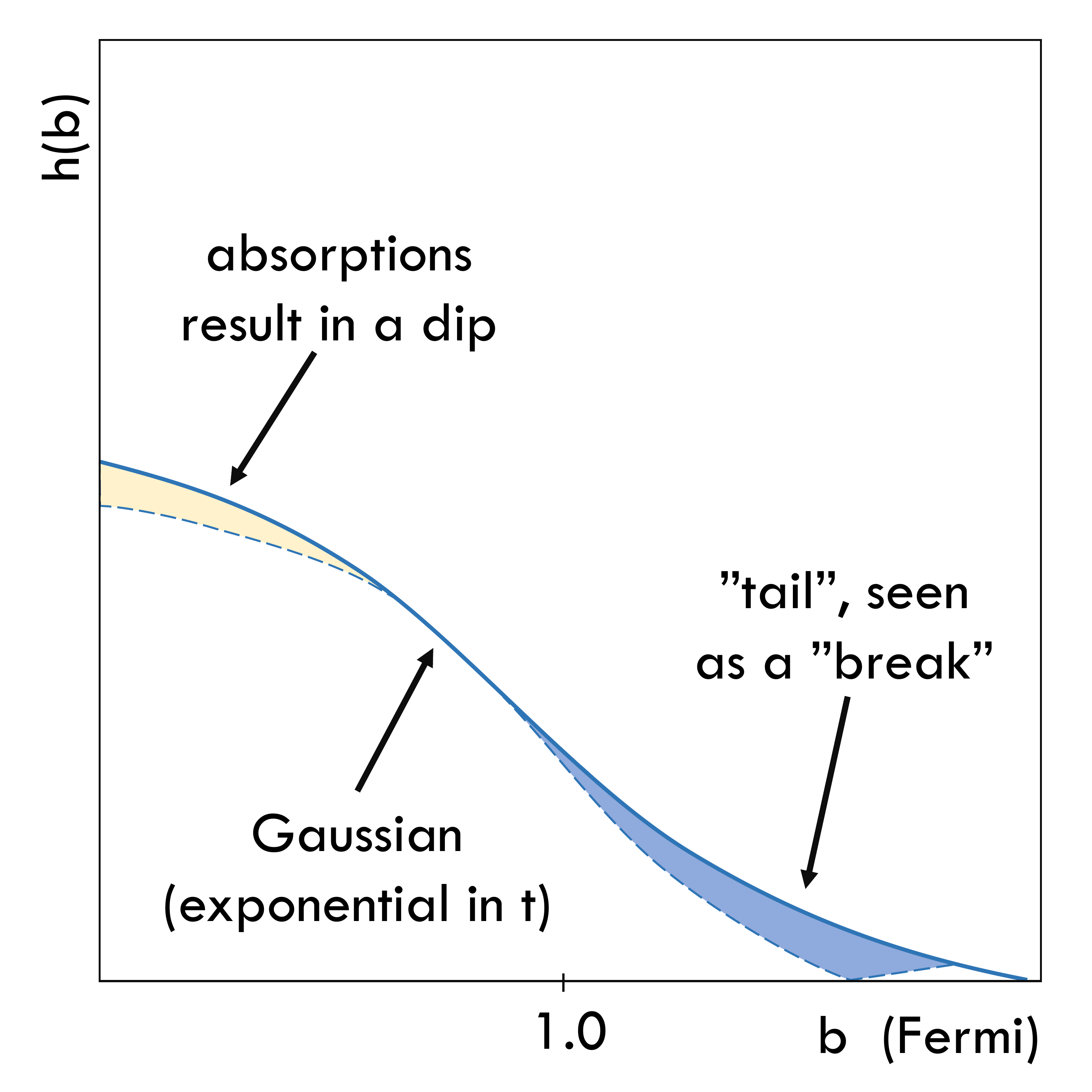}%
	}
	\caption{Schematic (qualitative) view of the "break", followed by the diffraction minimum ("dip"), shown both as function in $t$ and its Fourier transform (impact parameter representation), in $b$. While the "break" reflects the presence of the pion "atmosphere" (clouding) around the nucleon at peripheral values of $b$, the dip results from absorption corrections, suppressing the impact parameter amplitude at small $b$.}
	\label{Fig:1}
\end{figure}

\begin{figure}[h] 
\centering
\includegraphics[width=1\textwidth]{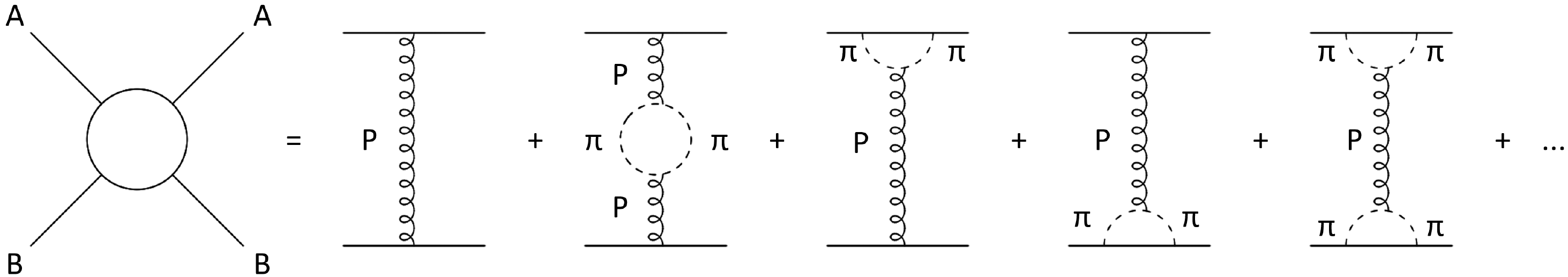}
\caption{Feynman diagram for elastic scattering with $t$-channel exchange containing a branch point at $t=4m_{\pi}^2$.} 
\label{Fig:Diagram}
\end{figure}

As illustrated in Fig.~\ref{Fig:1}, the "break" corresponds to the nucleon "atmosphere" (pion clouding). The effect, first observed at the ISR was interpreted as manifestation of $t$-channel unitarity, generating a two-pion loop in the cross channel (Fig.~\ref{Fig:Diagram}).    

The diffraction minimum instead is a consequence of $s$-channel unitarity (or absorption corrections) damping the impact-parameter amplitude at small-$b$, as shown in Fig.~\ref{Fig:1}.

\section{The "break" at the ISR and at the LHC} \label{s2}
In a recent paper \cite{RPM} the low-$|t|$ elastic $pp$ data, including the "break", was scrutinized at several energies. To answer the question about the universality of the fine structure of the diffraction cone (of the pomeron?!) we have extrapolated the low-$|t|$ structure from the ISR \cite{Bar} to the LHC \cite{TOTEM8}. To do so, a simple Regge-pole model was used, in which the "break" was parametrized by means of a non-linear pomeron trajectory with a two-pion threshold corresponding to the loop of Fig.~\ref{Fig:Diagram}. The scattering amplitude was constructed by a supercritical pomeron $A_P$ and an effective reggeon $A_f$ contributions:
\begin{equation} 
A(s,t)=A_P(s,t)+A_f(s,t),
\end{equation} 
where
\begin{eqnarray}
A_P(s,t)=-a_Pe^{b_P\alpha_P(t)}e^{-i\pi\alpha_P(t)/2}(s/s_0)^{\alpha_P(t)}, \nonumber \\ 
A_f(s,t)=-a_fe^{b_f\alpha_f(t)}e^{-i\pi\alpha_f(t)/2}(s/s_0)^{\alpha_f(t)}
\end{eqnarray}
with the trajectories
\begin{eqnarray}
\alpha_P(t)=\alpha_{0P}+\alpha'_Pt-\alpha_{1P}(\sqrt{4m_{\pi}^2-t}-2m_{\pi}), \nonumber \\  
\alpha_f(t)=\alpha_{0f}+\alpha'_ft-\alpha_{1f}(\sqrt{4m_{\pi}^2-t}-2m_{\pi}).
\end{eqnarray}
We use the norm:
\begin{equation}
\frac{d\sigma}{dt}(s,t)=\frac{\pi}{s^2}|A(s,t)|^2.
\end{equation}
Here $m_\pi$ is the pion mass, $4m_{\pi}^2=0.08$ GeV$^2$, $s_0=1$ GeV$^2$ and the values of fitted parameters are shown in Tab.~\ref{tab1}.

\begin{figure}[H] 
	\centering
	\subfloat[23.5 GeV\label{fig:R1}]{%
		\includegraphics[scale=0.25]{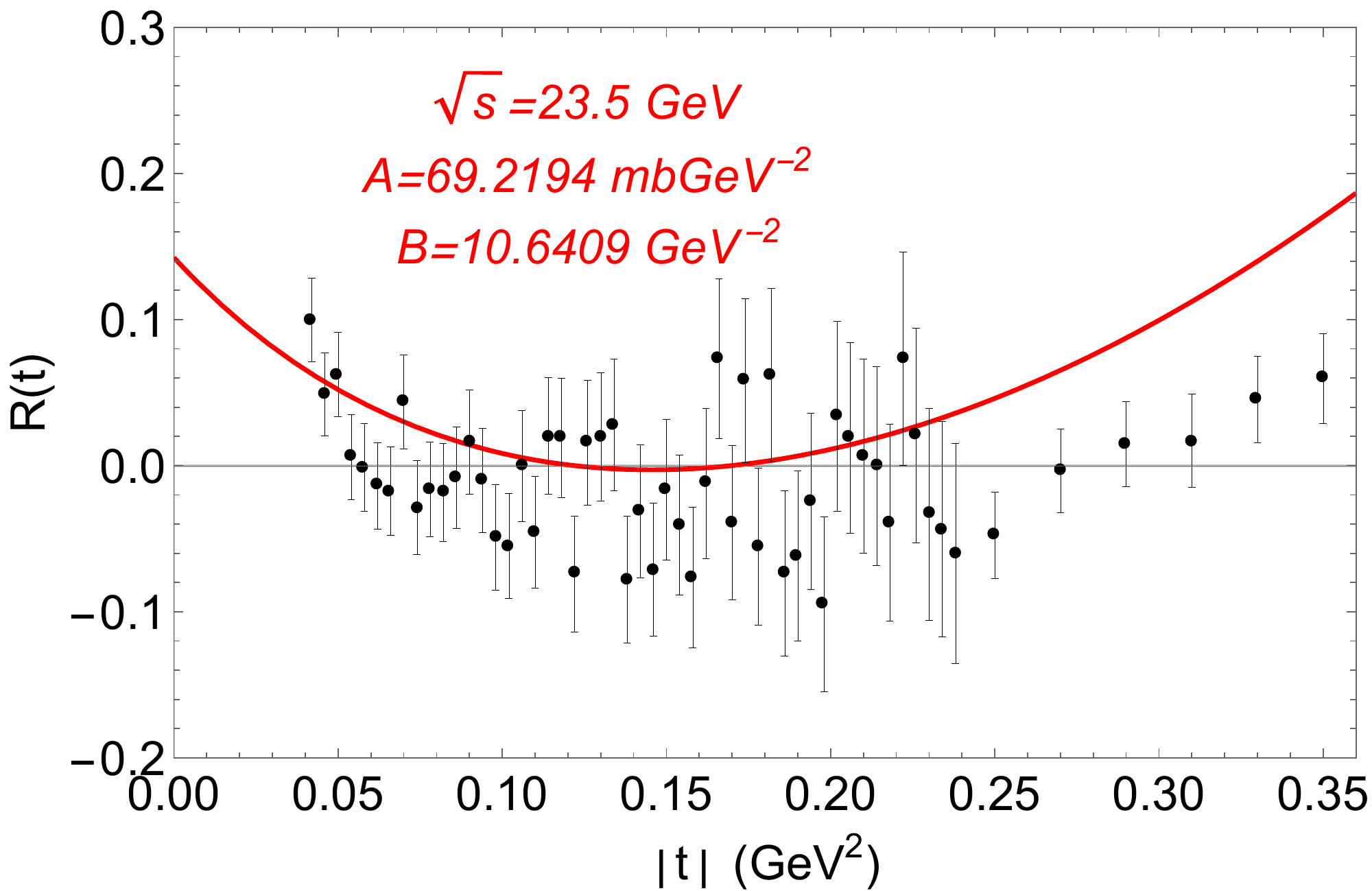}%
	}\hfill
	\subfloat[30.7 GeV\label{fig:R2}]{%
		\includegraphics[scale=0.25]{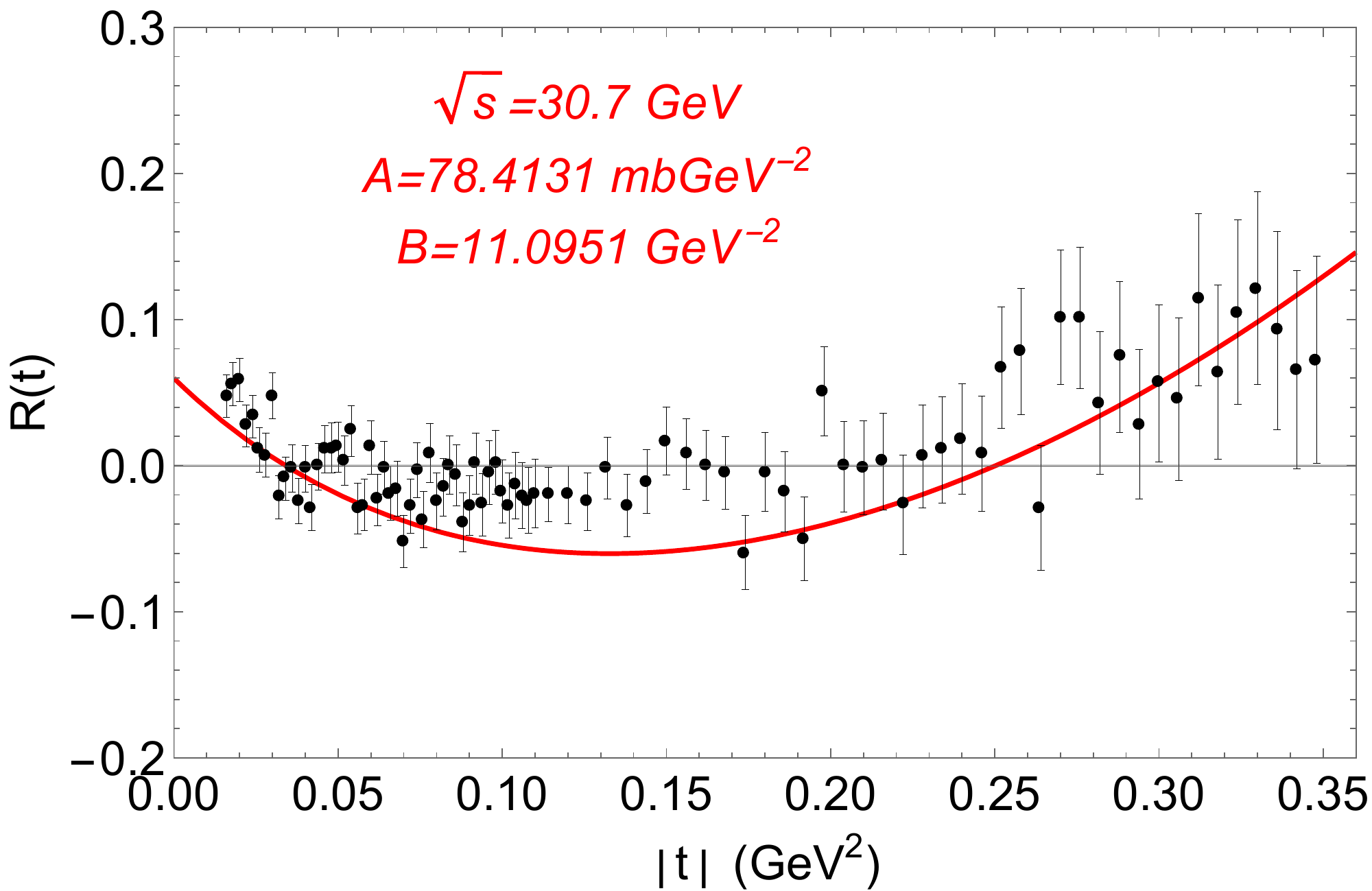}%
	}\hfill
	\subfloat[44.7 GeV\label{fig:R3}]{%
		\includegraphics[scale=0.25]{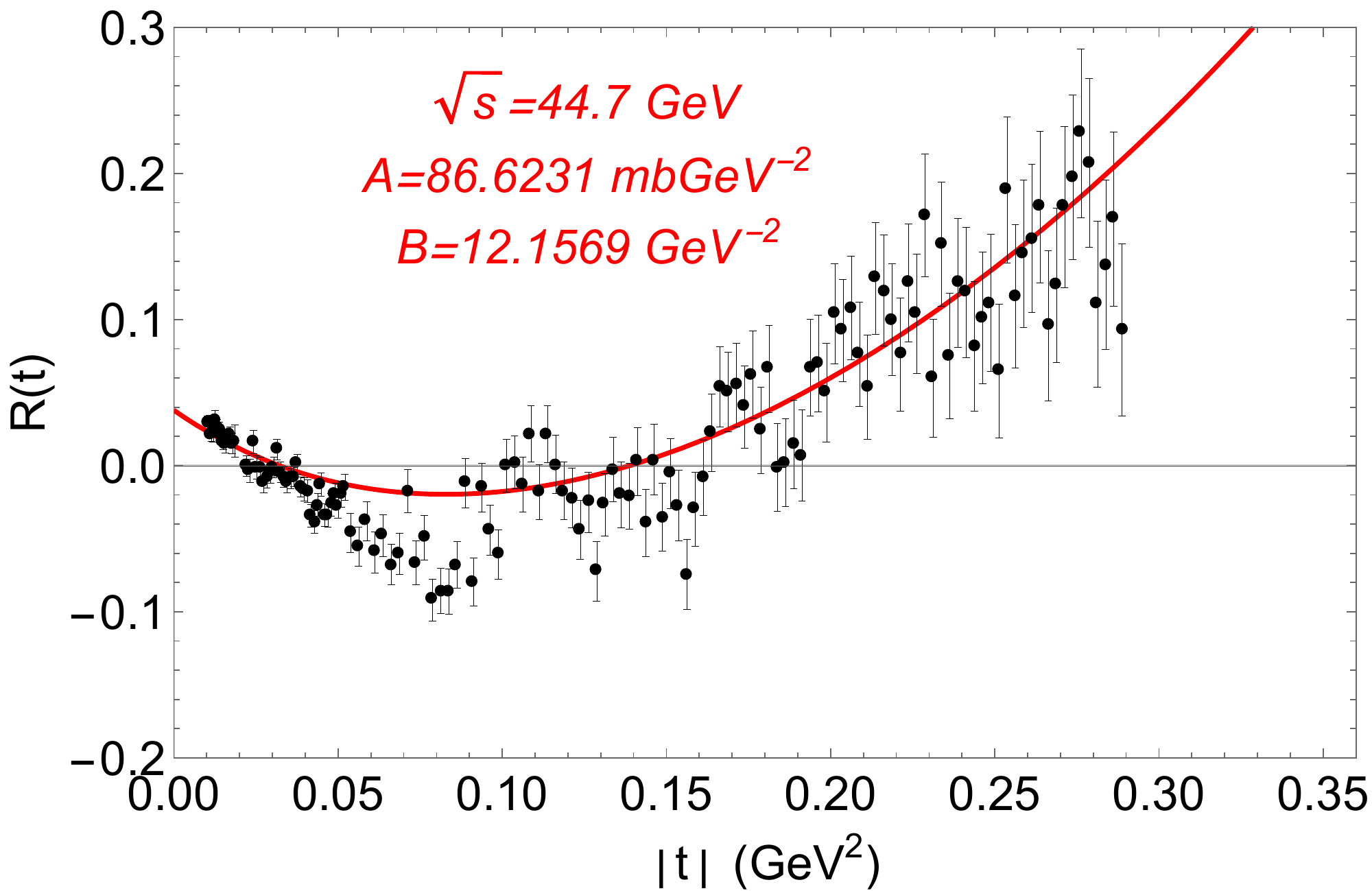}%
	}\hfill
	\subfloat[52.8 GeV\label{fig:R4}]{%
		\includegraphics[scale=0.25]{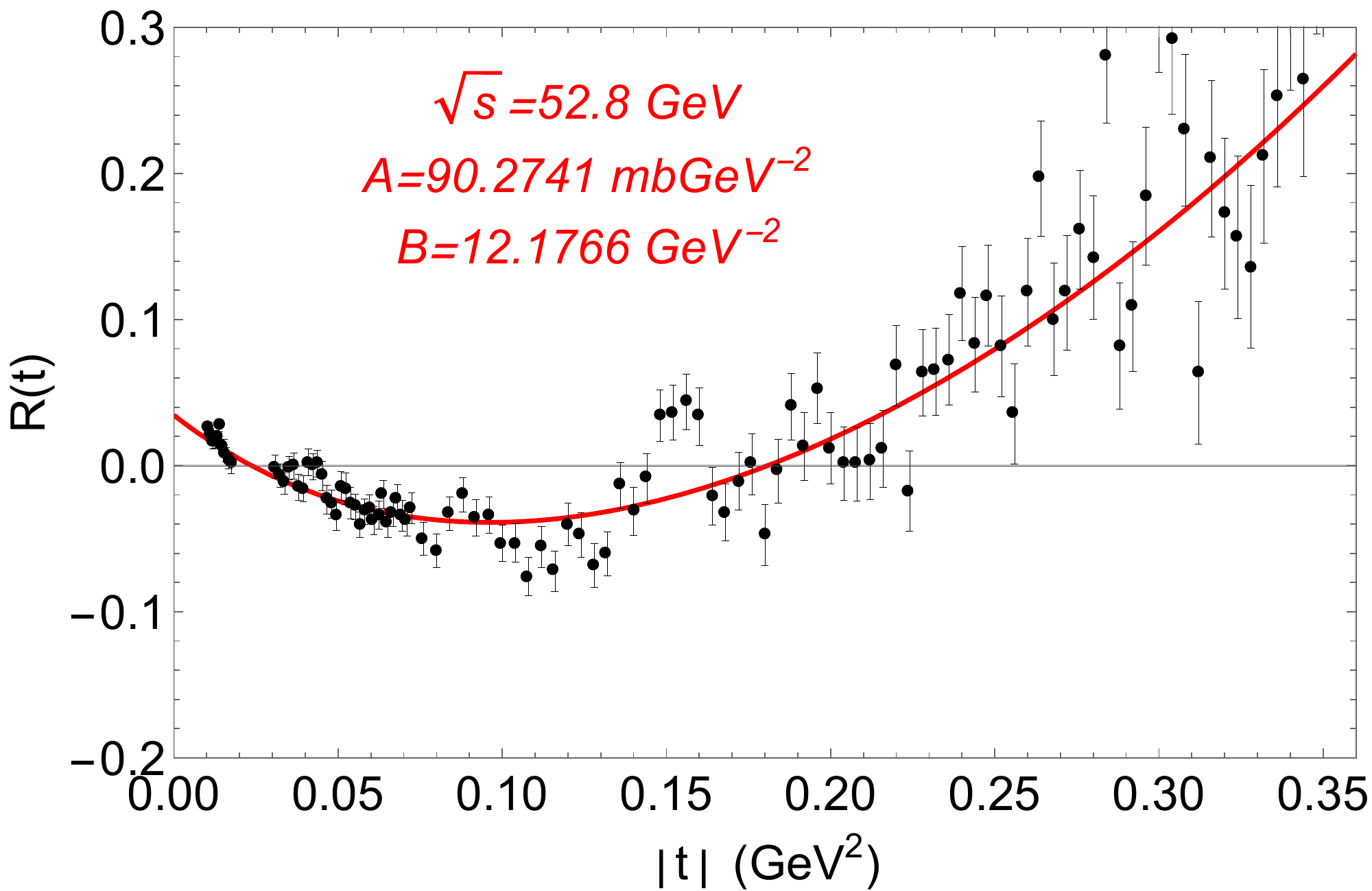}%
	}\hfill
	\subfloat[62.5 GeV\label{fig:R5}]{%
		\includegraphics[scale=0.25]{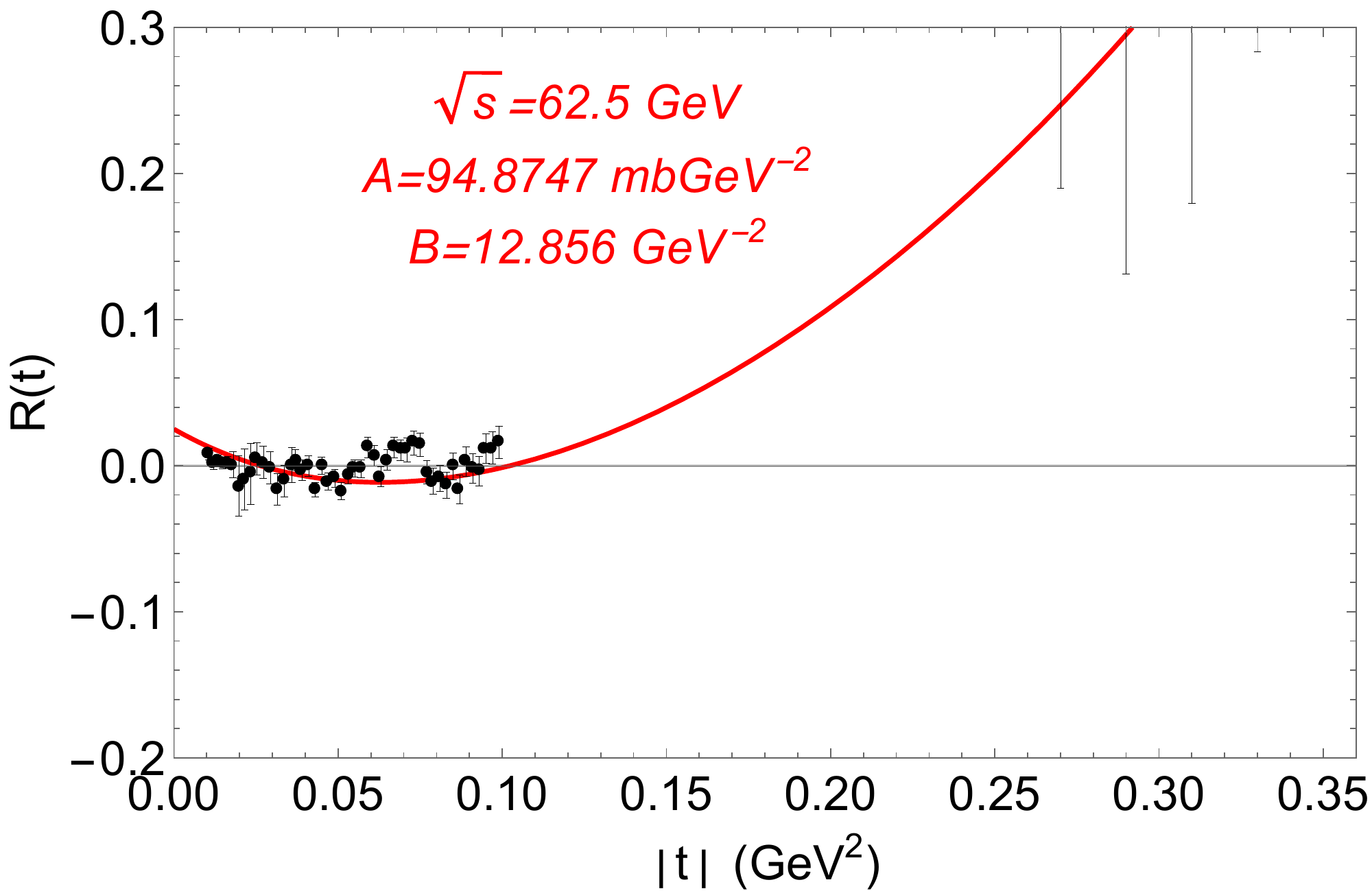}%
	}\hfill
	\subfloat[8 TeV\label{fig:R6}]{%
		\includegraphics[scale=0.25]{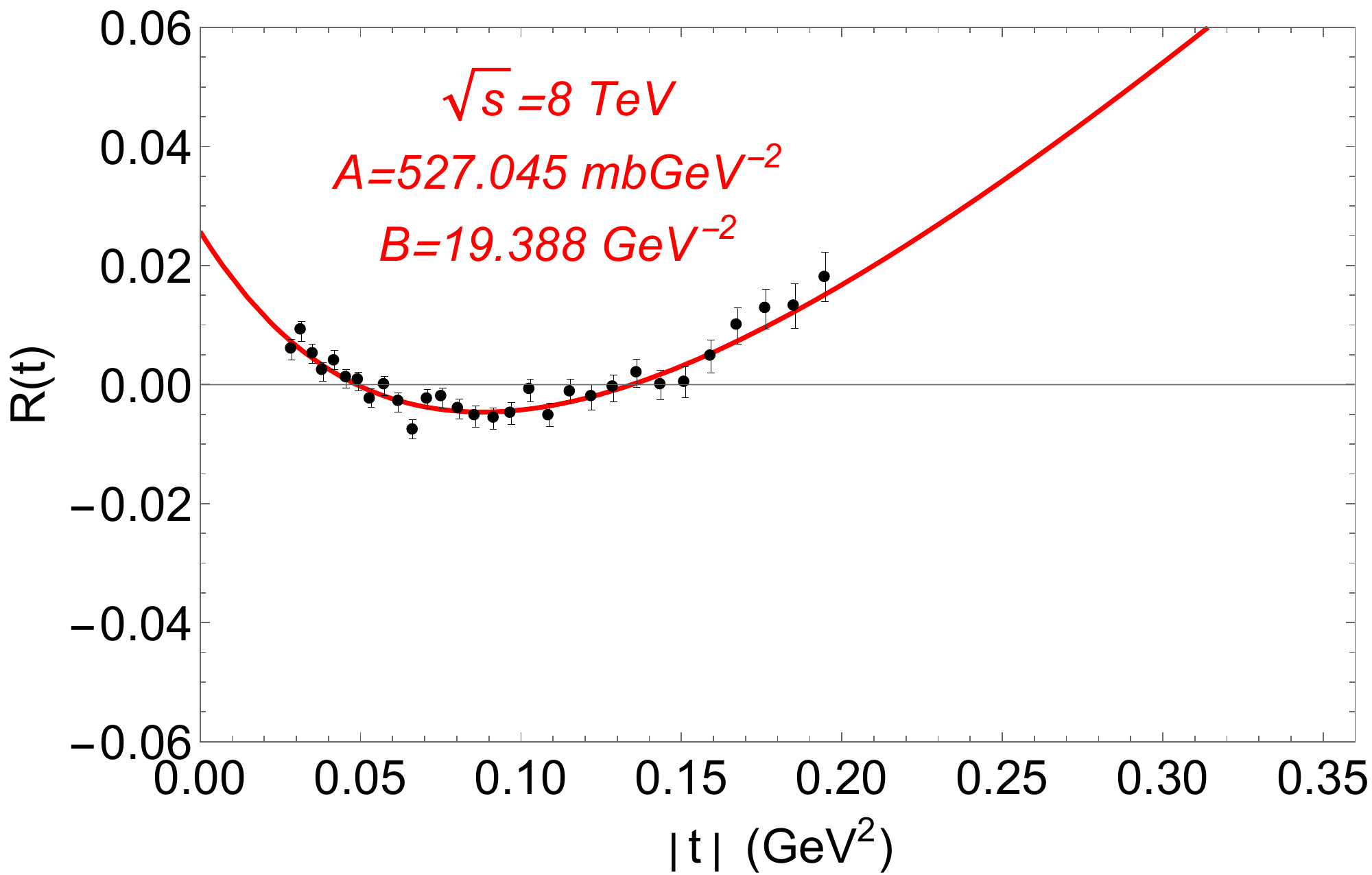}%
	}
	
	\caption{Calculated normalized differential cross sections -- results of mapping the ISR "break" to that at the LHC.}
	\label{Fig:Rratio}
\end{figure}

At the LHC the "break" is exposed by TOTEM \cite{TOTEM8} by means of the normalized differential cross section  
\begin{equation} \label{Eq:norm}
R=\frac{d\sigma/dt-ref}{ref},
\end{equation}
where $ref=Ae^{Bt}$ with $A$ and $B$ determined from a fit to the experimental data. The results of mapping the "break" through the energy from the ISR to the LHC are shown in Fig.~\ref{Fig:Rratio} in the normalized form.  

\begin{table}[tbph!]
	\begin{center}
		\begin{tabular}{|c|c|c|c|}
			\hline 
			\multicolumn{2}{|c|}{Pomeron} &  \multicolumn{2}{|c|}{Effective reggeon} \\
			\hline 
			$a_P$ & $0.0681012\pm0.0333978$  & $a_f$ & $0.18991\pm0.514455$ \\
			$b_P$ & $3.25783\pm0.568682$  & $b_f$ & $3.58861\pm2.89681$          \\
			$\alpha_{0P}$ & $1.11436\pm0.00591921$ & $\alpha_{0f}$ & $0.911087\pm0.0281833$ \\
			$\alpha'_{P}$ & $0.444551\pm0.0163735$  & $\alpha'_{f}$ & $0.887626\pm0.396745$ \\
			$\alpha_{1P}$ & $0.00246839\pm0.00798562$ & $\alpha_{1f}$ & $0.0777937\pm0.0668819$ \\ \hline 
			\multicolumn{4}{|c|}{$\chi^2/dof=2.2$} \\
			\hline
		\end{tabular}
	\end{center}
	\caption{Values of fitted parameters in mapping the ISR "break" onto that at the LHC.}
	\label{tab1}
\end{table}

\section{Correlation between the "break" and "dip"} \label{s4}

\begin{figure}[h] 
	\centering
	\includegraphics[width=.7\textwidth]{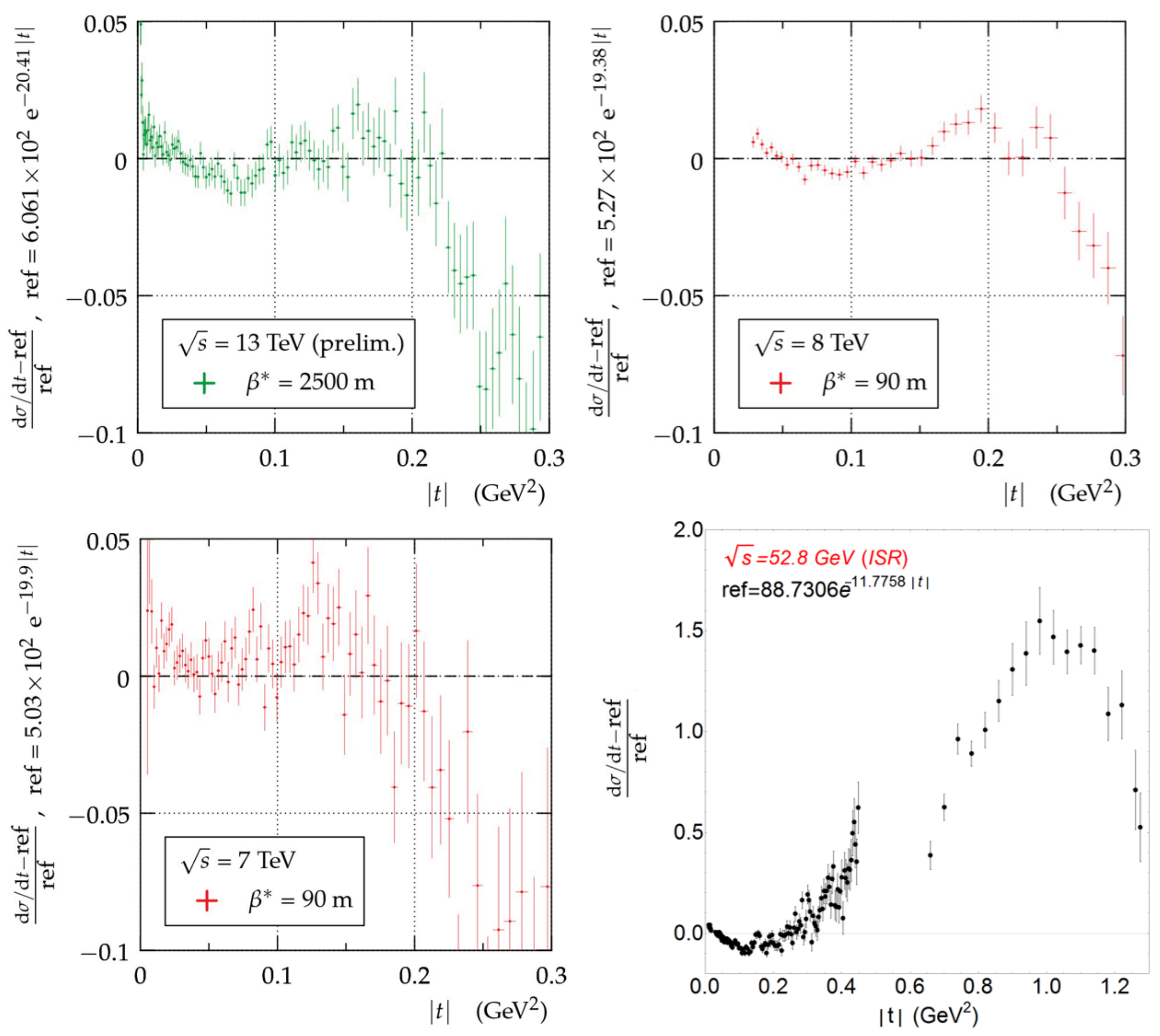}
	\caption{Effect of the "vortex" behavior of the "dip". Figures of TOTEM measurements at 7, 8 and 13 TeV are taken form Ref.~\cite{Deile}.} 
	\label{Fig:bd_correl}
\end{figure}

Fig.~\ref{Fig:bd_correl} shows that the normalized cross section calculated for TOTEM 7, 8 and 13~TeV data drastically deceases starting from $|t|\approx$ 0.2 GeV$^2$. Calculating the normalized cross section for the ISR 52.8 GeV data in their high exponential range (Fig.~\ref{Fig:bd_correl}) the same decrease can be observed from higher $|t|$ value. The reason of this phenomenon is the vicinity of the dip \cite{JSZ2}. The "dip" is acting as a "vortex" inclining the differential cross section to drop towards the dip position. 

To see clearly the "break", it is necessary to separate it from the influence of the neighbouring dip. The details of the "break" can be identified by optimizing the $t$-range studied/fitted with account of the correlations with the neighbouring dip.  

\section{Conclusions}
As emphasized at the beginning of the paper, the two structures on the otherwise exponential diffraction cone have quite different origin and physical meaning. Despite of it, as the "dip" comes towards smaller $|t|$ values it affects to the parametrization of the "break". It is important to take into account their correlation to identify the "break" at the LHC and higher energies.

The "break" as seen at the ISR and at the LHC at $\sqrt s=8$ TeV are of similar nature: they appear nearly at the same value of $t\approx -0.1$~GeV$^2$ and may be fitted by the same $t$-dependent function. 

Theoretical calculations of the relative contribution of the loop diagram, relative to the "Born term" (the ratio of two diagrams in the r.h.s. of Fig.~\ref{Fig:Diagram}), is of great importance, however relevant calculations are beyond the capability of perturbative~QCD. 

In a recent paper \cite{JSZT} we investigate the relative weight of two-pion effect to the vertex coupling (Regge residue) compared to expanding size (pomeron propagator) in producing the "break". We find that the effect primarily comes from the Regge residue (proton-pomeron coupling), rather than from the Regge propagator.

The "dip" region is also an important and complicated issue. Despite intense work along these lines (see e.g. Ref.~\cite{JLL} and earlier references therein), the existing models or theories are not able to predict unambiguously the details of this important phenomenon. At the dip, unlike the forward region, the odderon may become visible, moreover important, making different its shape in $pp$ and $\bar pp$ scattering, however increasing also the number of degrees of freedom. New data from the TOTEM Collaboration at the LHC at $13$ TeV will become public soon.



\Acknowledgements 
I would like to thank to the organizers of the EDS 2017 their support and the inspiring discussions at the Conference.

\end{document}

%% file: econfmacros.tex
\def\beq{\begin{equation}}
\def\eeq#1{\label{#1}\end{equation}}
\def\eeqn{\end{equation}}

\def\beqa{\begin{eqnarray}}
\def\eeqa#1{\label{#1}\end{eqnarray}}
\def\eeqan{\end{eqnarray}}

\let\bar=\overbar

\def\Dslash{\not{\hbox{\kern-4pt $D$}}}
\def\dslash{\not{\hbox{\kern-2pt $\del$}}}

\def\msb{{\bar{\ssstyle M \kern -1pt S}}}

